\documentstyle[aps]{revtex}

\begin{document}
\title{Virtual lattice dynamics method in quantum mechanics\bigskip\ }
\author{V.N.Pyrkov and V.M.Burlakov}
\address{Institute of Spectroscopy Russian Academy of Sciences, 142092 Troitsk,\\
Moscow region, Russia}
\maketitle

\begin{abstract}
\bigskip 

General molecular dynamic approach, making possible direct calculation of
eigen values and eigen functions for a quantum-mechanical system of an
arbitrary symmetry is proposed. The method is based on analogy between
discrete representation of the Schr\"{o}dinger equation and the system of
Newton equations describing dynamics of specially constructed virtual
lattice. Few examples demonstrating the method capabilities are considered.%
\bigskip\ 

PACS Numbers 73.20.At, 73.20.Hb\bigskip\ 
\end{abstract}

Molecular dynamics (MD) is a technique which is widely used for simulation
vibration of atoms and molecules, growth of a crystal or interface, the
interaction of an adatom and surface, and a wide range of other
time-dependent phenomena. In the MD technique, the many-body classical
equations of motion are solved as a function of time, and the physical
process can be studied in real time if the instantaneous forces acting on an
atom are given. The latter are obtained either using some devised potentials
to mimic the complicated character of the forces in realistic objects \cite
{forces} or from electronic-structure calculations (see, e.g. \cite{Drabold}
and references therein). In the {\it ab initio} MD calculations the
numerical integration of the Newton equations of motion for a given number
of ions involves the forces derived, at each time step, from the
instantaneous electronic configuration.

While considerable efforts have been made to develop the quantum-mechanical
electronic structure calculations and to use the obtained Hellmann-Feynmann
forces in the MD simulations (see, e.g. \cite{Sankey}), little attention has
yet been given to applying the MD technique in the quantum mechanics
directly. Recently a fictitious Newtonian dynamics was introduced for
electronic variables to make the {\it ab-initio} MD calculations more
effective \cite{Car-par}. The physics behind this method was analyzed in 
\cite{Pastore}.

We present here a general MD approach, which allows direct calculation of
eigen values and eigen functions of any stationary quantum-mechanical
problem. To determine by ordinary methods the eigen functions are assumed to
be expanded onto a complete set of orthogonal functions. These can be either
atomic orbitals or plane waves depending on whether the localized or
delocalized states are described. Our method can be effectively used
regardless the localization properties of the studied states, geometry and
symmetry of a system. We restrict our consideration here by a one-particle
multicenter problem although the proposed method can in principle be applied
to many-particle problems as well.

The method is based on discrete representation of the Schr\"{o}dinger
equation for the particle of mass $m_e$ in the potential $U(x,y,z)$ 
\begin{equation}
-\frac{\hbar ^2}{2m_e}\cdot \left\{ \frac{\partial ^2\Psi (x,y,z)}{\partial
x^2}+\frac{\partial ^2\Psi (x,y,z)}{\partial y^2}+\frac{\partial ^2\Psi
(x,y,z)}{\partial z^2}\right\} +\left( U(x,y,z)-E\right) \cdot \Psi
(x,y,z)=0,\   \label{eq1}
\end{equation}
by a system of oscillator equations. Indeed\bigskip
\begin{eqnarray*}
\frac{\partial ^2\Psi (x,y,z)}{\partial x^2} &=&\frac{\Psi (x+\delta
_x,y,z)+\Psi (x-\delta _x,y,z)-2\Psi (x,y,z)}{\delta _x^2}+ \\
O(\delta _x^2) &\approx &\frac{\Psi (x+\delta _x,y,z)+\Psi (x-\delta
_x,y,z)-2\Psi (x,y,z)}{\delta _x^2}
\end{eqnarray*}
\ and analogously for the $\frac{\partial ^2\Psi (x,y,z)}{\partial y^2}$ and 
$\frac{\partial ^2\Psi (x,y,z)}{\partial z^2}$. Let the $\Psi $-function be
defined on discrete space, i.e. $x=n\cdot \delta _x$, $y=m\cdot \delta _y$, $%
z=l\cdot \delta _z$. Then 
\begin{eqnarray*}
\Psi _{n,m,l} &=&\Psi (n\cdot \delta _x,m\cdot \delta _y,l\cdot \delta _z),~
\\
U_{n,m,l} &=&U(n\cdot \delta _x,m\cdot \delta _y,l\cdot \delta _z)
\end{eqnarray*}
and Eq. (\ref{eq1}) can be approximated as\bigskip
\begin{equation}
\begin{array}{c}
-E\cdot \Psi _{n,m,l}+K_x\cdot \left( 2\Psi _{n,m,l}-\Psi _{n+1,m,l}-\Psi
_{n-1,m,l}\right) +K_y\cdot \left( 2\Psi _{n,m,l}-\Psi _{n,m+1,l}-\Psi
_{n,m-1,l}\right) + \\ 
K_z\cdot \left( 2\Psi _{n,m,l}-\Psi _{n,m,l+1}-\Psi _{n,m,l-1}\right)
+U_{n,m,l}\cdot \Psi _{n,m,l}=0,
\end{array}
\label{eq01}
\end{equation}
\ where $K_i=\hbar ^2/(2m_e\delta _i^2)\ (i=x,y,z)$. Suggesting 
\begin{equation}
\Psi _{n,m,l}(t)=\Psi _{n,m,l}\cdot \exp (i\sqrt{E}\cdot t)\ \   \label{eq1a}
\end{equation}
we can rewrite (\ref{eq01}) and obtain the system of linear equations 
\begin{equation}
\begin{array}{c}
\frac{d^2\Psi _{n,m,l}(t)}{dt^2}+K_x\cdot \left( 2\Psi _{n,m,l}(t)-\Psi
_{n+1,m,l}(t)-\Psi _{n-1,m,l}(t)\right) +K_y\cdot \left( 2\Psi
_{n,m,l}(t)-\Psi _{n,m+1,l}(t)-\Psi _{n,m-1,l}(t)\right) + \\ 
K_z\cdot \left( 2\Psi _{n,m,l}(t)-\Psi _{n,m,l+1}(t)-\Psi
_{n,m,l-1}(t)\right) +U_{n,m,l}\cdot \Psi _{n,m,l}(t)=0\ 
\end{array}
\label{eq2}
\end{equation}
describing the harmonic vibrations of the unity-mass virtual particles in a
lattice with nearest neighbor force constant $K_i=\hbar ^2/(2m_e\delta _i^2)$%
\ and incite force constant $U_{n,m,l}$. The $\Psi _{n,m,l}(t)$ in this
context means the displacement of the virtual particle in the cite ($n,m,l$)
from its equilibrium position. Thus, from continues equation (\ref{eq1}) for
the one-particle $\Psi $ function we have arrived to the system of Newton
equations (\ref{eq2}) describing the dynamics of the virtual lattice and the
eigen functions of the quantum-mechanical problem are associated now with
the eigen vectors of the virtual lattice. Therefore the described method one
may call as the {\it virtual lattice dynamics} (VLD) method.

For practical use of the VLD method one should construct first the virtual
lattice and then follow the usual way of MD simulation of lattice dynamics.
An important point concerns an {\it a priori} estimation of the accuracy of
the VLD method since it results in the proper choice of spacings $\delta _i$%
. The accuracy can be clearly controlled using the definition $%
E_j=\left\langle \Psi _j\left| \stackrel{\symbol{94}}{H}\right| \Psi
_j\right\rangle $. For the ground state $E_0$ one may neglect the kinetic
energy operator in the Hamiltonian keeping the potential term $U(x,y,z)$
only. As the $\Psi $ function is a smoother function than the potential one
and its discreteness can be neglected, the accuracy can be estimated on a
bases of approximation of the potential curve $U(x,y,z)$ by the discrete
function $U_{n,m,l}$. For 1-D case one may accept $\Psi _j=Const$ over the $%
\Psi _j$ function localization length $L_j=N_j\cdot \delta $ and then
roughly estimate the relative accuracy of the VLD method as $\sim 1/N_j$,
where $N_j$ is the number of the virtual particles inside the $L_j$. Similar
arguments can also be applied to a 3-D case where relative accuracy roughly
corresponds to the ratio $\sim N_j^S/N_j^V$, $N_j^S$ and $N_j^V$ being the
numbers of virtual particles on the surface of the $\Psi _j$ function
localization region and inside it respectively. Thus, the accuracy of the
VLD method obviously increases with decrease of spacing $\delta $ getting
asymptotically exact.

For illustration we consider the simple analytically solvable example from
quantum mechanics textbook: single unity-mass particle in semi-infinite 1-D
rectangular potential well (see Fig.1). The virtual lattice is represented
by a simple linear chain of particles separated by $\delta $ with
interparticle coupling constant $K=1/(2\delta ^2)$ ($\hbar =1$) and incite
force constant $U_n=0$ for $n<24$ and $U_n=0.1$ otherwise. We have chosen
the number of virtual particles inside the potential well $N_w=23$ ($\delta
=1$) and hence the {\it a priori} estimated accuracy in the determination of
eigen values must be about 4 per cent. At the $t=0$ the particles in the
chain were randomly displaced and their subsequent vibrations were analyzed
via Fourier transformation. According to (\ref{eq1a}) the eigen energy $E_j$
is given by $\omega _j^2$, where $\omega _j$ is the peak position in the
Fourier spectrum $\sum\limits_n\left| \Psi _n(\omega )\right| $. The Fourier
spectrum $\sum\limits_n\left| \Psi _n(\omega ^2)\right| $ shown in Fig.1
reveals five fairly separated sharp peaks corresponding to the bound states
within the potential well, and the continuum corresponding to extended
states. The calculated eigen values for the bound states deviate from those
obtained analytically by less than 2 per cent giving evidence for quite
reasonable {\it a priori} estimation of the accuracy. In case of $N_w=48$ ($%
\delta =1/2$) the deviation decreases down to $\sim $1 per cent and
obviously will tend to zero if $\delta \longrightarrow 0$.

After determination of the eigen values one can calculate the corresponding
eigen functions. In nondegenerate case the eigen function $\Psi _j(x)$ can
be approximated as 
\begin{equation}
\Psi (x)\simeq \Psi (n\cdot \delta )=C^{-1}\cdot 
\mathop{\rm Re}
\left[ \Psi _n(\omega _j)\right] ,  \label{eq3}
\end{equation}
where $C=\sum\limits_{n=1}^N\left\{ 
\mathop{\rm Re}
\left[ \Psi _n(\omega _j)\right] \right\} ^2$ is normalization factor, $N$
is the total number of virtual particles in the chain. Another and more
exact though more time consuming way to determine the $\Psi _j(x)$ is as
follows: a virtual particle at a nonsymmetric position $m$ is initially
excited by a harmonic force with eigen frequency $\omega _j=\sqrt{E_j}$. To
reach the steady state condition of the excitation the small
phenomenological damping $\gamma \cdot \frac{\partial \Psi _n(t)}{\partial t}
$ was introduced into the motion equations (\ref{eq2}). After few tens of
vibrations the spatial character of the virtual particles vibration around $%
m-th$ particle is recorded and the excitation force field is made identical
to the particles displacement field. The procedure is repeated until no
noticeable difference is observed between the excitation and the particles
displacement fields. The final spatial character of the excitation must be
identical to the corresponding $\Psi _j(x)$ function since no other states
can be excited if there is no degeneracy. Thus obtained $\Psi _j(x)$
functions (see, e.g. Fig.1) are in perfect agreement with the theoretical
ones.

To check if the eigen state is degenerated a different virtual particle have
to be chosen for the initial excitation with the same frequency $\omega _j$.
The state is nondegenerated if the new determined $\Psi _j(x)$ function
coincides with that determined previously for the $m-th$ virtual particle as
the initially excited one. Otherwise one should determine the degree of
degeneracy by taking different virtual particles for the initial excitation
and then construct the orthogonal combinations from all determined $\Psi
_j(x)$ functions for the given $\omega _j$.

To illustrate the applicability of the VLD method to a more complicated
system we calculated the energy levels and the charge density distribution
in the bulk and near the surface of the 2-D two-atomic crystal band with
impurities (Fig.2). As an ordinary MD technique the VLD method is restricted
by the number of particles under treatment. Being able to handle effectively
about a billion of virtual particles we describe the crystal band consisting
of $10\times 40$ potential wells (atoms) with an {\it a priori }estimated
accuracy of few per cent. Virtual lattice in this case is a simple quadratic
one extended outside the crystal band in the x-direction. Cyclic boundary
conditions in both x- and y-directions for the virtual lattice vibrations
were applied. Panels (a) and (b) in Fig.3 show the shape of the atomic
potential. The potential parameters were chosen in a way to form a surface
electronic band near the middle of the band gap in the energy spectrum. The
latter obtained via Fourier analysis of the virtual lattice vibrations is
shown in Fig.3c. The potential of impurity atoms (denoted by stars in Fig.3)
was chosen to be only slightly different from that of one of the host atoms.

Using the excitation in the form of standing wave with different period the
dispersion of the electronic excitations over the Brillouin zone was
obtained (Fig. 4). Different dispersion curves in Fig.4 can be associated
with electronic bands of different symmetry, what is not a subject of our
paper, however. We are interested basically in discrimination of the states
with regard to the wave vector $k_y$.

Spatial charge distribution $\rho _j(x,y)\sim \Psi _j^2(x,y)$ in different
eigen states corresponding to the Brillouin zone centre $k_y=0$ are shown in
Fig.5 for both pure and impurity crystals. The potential defect related to
impurity atom could not result in the localization in the bulk of the
crystal, but turned out to be big enough to form the localized states at the
surface (see Fig.6). The latter are strongly enhanced around the impurity
clusters near the surface. The VLD method allows also to study an influence
of surface roughness on electronic states. Charge distribution in the bulk
eigen states disturbed by the surface roughness is shown in Fig.7. The
roughness induced localized state at the surface is presented in Fig.8.

The considered example showed that the VLD method can be effectively used
for numerical studies of one-particle eigen states in inhomogeneous and low
symmetry systems. The method seems quite adequate for direct investigation
of defect and/or impurity states in compensated semiconductors when the
interaction between one-particle localized states of donors can be
neglected. In case of the nonvanishing interaction the VLD method can still
be applied via proper iteration procedure. Worth to note one important
feature of the VLD method: remaining asymptotically exact it can be
generalized on the case of few particles in multicenter potential allowing
to solve excitonic problem in a cluster of an arbitrary symmetry.

The work was partially supported by Russian Ministry of Science within the
program ''Fundamental Spectroscopy''.

\begin{center}
\newpage\ Figure captures
\end{center}

\begin{description}
\item  Fig.1. One-particle states in the semi-infinite rectangular potential
well (shown by thick solid lines in the right panel) $U_n=0$ if $n\leq 38$
and $U_n=0.2$ otherwise. The energy levels correspond to the peak positions
in the Fourier spectrum $\sum \left| \Psi _n(\omega ^2)\right| $ (right
panel). The eigen functions calculated for three lowest lying eigen states
are shown in the left panel.

\item  Fig.2. Two atomic 2-D crystal with impurities. Axes labels are given
in $\delta $ units thus corresponding to the virtual particle number. Large
and small black points denote different potential wells corresponding to two
different types of the host atoms. The atoms with deeper potential are
randomly substituted by impurity atoms (white points). Atomic potential
profile is shown in Fig.3.

\item  Fig.3. Potential profile of the crystal shown in Fig.2: a) along
x-axes at y=6; b) along y-axes at x=133; c) Fourier spectrum of the virtual
lattice vibration showing the valence (V) and conduction (C) bands in the
bulk, surface band (S), and localized states (L) due to impurity atoms at
the surface.

\item  Fig.4. Dispersion of eigen energies over the Brillouin zone for the
2-D two atomic crystal a) pure; b) with impurities (Fig.2); c) pure crystal
with rough surface (Fig.7a). Surface band is labelled by S. Open circles
denote eigen energies for which the eigen functions were calculated (Figs 5
- 8).

\item  Fig.5. Charge density distribution $\rho _j(x,y)\sim \Psi _j^2(x,y)$
calculated for some eigen states of the pure ((a), (b), and (c)) and the
impurity ((d), (e), and (f)) crystals: (a) and (d) correspond to the state
1, (b) and (e) - 2, (c) and (f) -3, denoted in Fig.4. The states (a), (b),
(d), and (e) are obviously the bulk ones while (c) and (f) are the surface
ones.

\item  Fig.6. Charge density distribution $\rho _j(x,y)\sim \Psi _j^2(x,y)$
in the localized surface state (right panel) of the impurity crystal
(labelled as IL in Fig.4) in comparisom with the extended surface state
(left panel) of the pure crystal (3 in Fig.4). Open circles in the right
panel denote impurity atoms.

\item  Fig.7. $\Psi _j^2(x,y)$ calculated for the pure crystal with rough
surface (a). Panels (b), (c), (d) correspond respectively to the states 1,
2, 3 in Fig.4.

\item  Fig.8. $\Psi _j^2(x,y)$ calculated for the roughness induced
localized (b) surface state (RL in Fig.4) in comparison with the extended
surface state (a) of the pure crystal (4 in Fig.4).
\end{description}

\end{document}